# Micro-structure and Magnetization of the 80-K Superconductor, TbSr$_2$Cu$_{2.7}$Mo$_{0.3}$O$_{7+\delta}$

V.P.S. Awana[1], Anurag Gupta[1], H. Kishan[1], M. Karppinen[2], H. Yamauchi[2], A.V. Narlikar[3,*], E. Galstyan[4] and I. Felner[4]

[1]National Physical Laboratory K.S. Krishnan Marg, New Delhi – 110012, India.
[2]Materials and Structures Laboratory, Tokyo Institute of Technology, 226-8503, Yokohama, Japan.
[3]Inter-University Consortium for DAE Facilities, University Campus, Khandwa Road, Indore-452017, MP, India.
[4]Racah Institute of Physics, The Hebrew University, Jerusalem, 91904, Israel

The results of micro-structural and detailed magnetization studies are reported here for our recently published TbSr$_2$Cu$_{2.7}$Mo$_{0.3}$O$_{7+\delta}$ superconductor with a transition temperature ($T_c$) at as high as 80 K [1]. From XRD and EDX the sample was confirmed to be of single-phase and possess the nominal cation stoichiometry. The SEM images showed well-developed grains for this ceramic high-$T_c$ superconductor. Magnetization measurements at various temperatures and fields revealed bulk superconductivity below 80 K. Magnetization versus applied field (M vs. H) loops at various temperatures below the $T_c$ exhibited clear full penetration field $H_P$. In normal state (> 100 K) the susceptibility follows the Curie-Weiss behavior with an effective paramagnetic moment of 9.83 $\mu_B$.

**Key words:**

Micro-structure, SEM, EDX, Bulk superconductivity, Magnetization, and TbSr$_2$Cu$_{2.7}$Mo$_{0.3}$O$_7$ compound

\* **Presently at:**

Racah Institute of Physics, The Hebrew University, Jerusalem, 91904, Israel

## 1. INTRODUCTION

Most of the high-$T_c$ superconductors (HTSCs) discovered since 1993 have been synthesized through high-pressure high-temperature synthesis routes [2]. Application of ultra high pressures (2 ~ 8 GPa) at higher temperatures (>1000 $^0$C) (HPHT-syn) has contributed in realizing various new superconducting layered perovskite structures [3]. Loading an ambient-pressure synthesized copper-oxide phase with excess oxygen is another powerful option provided by the high-pressure techniques. Here the pre-synthesized sample is treated under high pressures but at somewhat lower temperatures (400 ~ 600 $^o$C) in the presence of additional oxygen-releasing oxide, such as AgO, KClO$_4$, etc. [3], i.e. HPLT-oxy. One such example was reported by some of us very recently [1], regarding the enhancement of the superconducting transition temperature ($T_c$) of the TbSr$_2$Cu$_{2.7}$Mo$_{0.3}$O$_{7+\delta}$ phase up to 80 K by a HPLT-oxy process. As discussed in ref. [1], there are various problems in realizing the "90-K phase" of TbBa$_2$Cu$_3$O$_7$ or CuBa$_2$TbCu$_2$O$_7$ (Tb-123 or Cu-1$^{(Ba)}$2$^{(Tb)}$12), like the formation of stable TbBaO$_3$ rather than the desired Tb-123 phase [4,5]. To overcome this situation, Sr is used in place of Ba and subsequently to stabilize the structure, approximately 30 % of the Cu atoms in CuO chains is replaced by Mo. This resulted in the composition of TbSr$_2$Cu$_{2.7}$Mo$_{0.3}$O$_{7+\delta}$, which not only crystallizes in ideal Tb-123 or (Cu,Mo)-1$^{(Sr)}$2$^{(Tb)}$12 structure, but also shows superconducting transition temperature $T_c$ of up to 37 K [6-9]. What we reported earlier [1], is the enhancement of the $T_c$ of the TbSr$_2$Cu$_{2.7}$Mo$_{0.3}$O$_{7+\delta}$ phase from 37 K to ~80 K by a HPLT-oxy process (5 GPa, 400 $^0$C) using AgO as an excess-oxygen source.

In HTSC compounds increasing the oxygen content provides increased numbers of mobile hole carriers, which might result in higher $T_c$ values. Interestingly though HPHT-syn and HPLT-oxy treatments have been very useful in realizing new or improved HTSC systems, the same has various disadvantages also, viz. micro-cracks, dislocations, compositional variations and non-connectivity of the relatively small grains. Unlike for normal pressure heat treatments of several hours with intermediate mixing/grindings, in HPHT-syn and HPLT-oxy processes the compound is treated under ultra high pressures for a few tens of minutes only. This sort of "shock treatment" may give rise to various problems at micro-level in the end product.

In the present paper we report the results of micro-structural and detailed magnetization studies on our HPLT oxygenated $TbSr_2Cu_{2.7}Mo_{0.3}O_{7+\delta}$ compound with $T_c$ of 80 K. Micro structural analysis confirmed that our single-phase $TbSr_2Cu_{2.7}Mo_{0.3}O_{7+\delta}$ sample is composed of well-developed grains with cation stoichiometry that is essentially the same as the nominal one. Magnetization measurements at various temperatures and fields showed the compound to be bulk superconducting at 80 K. Magnetization versus applied field (M vs. H) loops at various temperatures below $T_c$ exhibited clear full penetration field ($H_P$) values.

## 2. EXPERIMENTAL DETAILS

The 80-K superconducting $TbSr_2Cu_{2.7}Mo_{0.3}O_{7+\delta}$ sample was synthesized by a two-step process, as described in detail in ref. [1]. In short, the $TbSr_2Cu_{2.7}Mo_{0.3}O_{7+\delta}$ phase with $\delta = 0.12$ was first obtained by through solid-state synthesis in air. The further oxygenation ($\delta > 0.12$) was facilitated through a HPLT-oxy process by annealing the as-air-synthesized sample in a cubic-anvil-type high-pressure apparatus at 5 GPa and 400 $^o$C for 30 min in the presence of additional AgO (100 mol-%) as an excess-oxygen source. An X-ray powder diffraction pattern was obtained by a diffractometer (Philips-PW1800) with Cu $K_\alpha$ radiation. The microstructure and the phase integrity of the sample were investigated by QUANTA (Fri Company) scanning electron microscopy (SEM) and by a Genesis energy dispersive x-ray (EDX) analysis device attached to the SEM. DC susceptibility data were collected by a SQUID magnetometer (Quantum Design, MPMS).

## 3. RESULTS AND DISCUSSION

Figure 1 shows a typical SEM picture of our HPLT-oxy $TbSr_2Cu_{2.7}Mo_{0.3}O_{7+\delta}$ sample. It is evident that the sample has a granular structure. In the picture small 1-4 $\mu$m sized well-sintered grains are seen. The grains were concluded to be of the single Tb-123 or $(Cu,Mo)-1^{(Sr)}2^{(Tb)}12$ phase. Using EDX, compositional analysis was carried out on several grains shown in Fig. 1. The analysis yielded a cation stoichiometry of 1.0(1)Tb-1.9(1)Sr-2.5(1)Cu-0.5(1)Mo that is (within the uncertainty limit) very close to the nominal stoichiometry of the $TbSr_2Cu_{2.7}Mo_{0.3}O_{7+\delta}$ compound. We may thus conclude that during our two-step synthesis route no cation evaporation occurred. Note that among the constituent metals, Mo is expected to be the one that most easily might evaporate; Here we rather observed a slightly Mo-rich composition completely ruling out the possible evaporation of it.

Fig. 2(a) depicts the DC susceptibility versus temperature plots for HPLT-oxy TbSr$_2$Cu$_{2.7}$Mo$_{0.3}$O$_{7+\delta}$ compound under an applied field of 6 Oe [ZFC (zero-field-cooled) and FC (field-cooled) modes]. Four characteristic temperatures namely $T_c$, $T_{irr}$, $T_K$ and $T_N$ are seen. Here $T_c$ (80 K) is the superconducting transition temperature and is defined as the deviation of susceptibility from normal state to superconducting diamagnetic state. $T_{irr}$ (75 K) is the irreversibility temperature where ZFC and FC susceptibility branches merge. $T_K$ (35 K) seems to be due to some magnetic ordering inherent in the compound. The fourth characteristic temperature $T_N$ (7 K) is the antiferromagnetic ordering temperature of Tb moments [1,6-9] Interestingly $T_N$ of Tb in Tb-123 remains unchanged irrespective of whether the compound is non-superconductiing or superconducting at elevated temperatures of 30 or 80 K. Hence out of the four characteristic temperatures seen in DC susceptibility though the origin of $T_c$, $T_{irr}$ and $T_N$ is well understood, the origin of $T_K$ (35 K) needs to be debated. Interestingly $T_K$ appears only for HPLT-oxy TbSr$_2$Cu$_{2.7}$Mo$_{0.3}$O$_{7+\delta}$, and is not seen for 100-atm O$_2$ annealed or for the as synthesized compound [1]. It may be that $T_K$ is related to a small fraction of HPLT-oxy synthesized TbCuO$_{2+d}$ delafossite phase. Magnetism of these phases is known to be of quasi two-dimensional in nature having strong dependence on applied fields [10].

Keeping this in mind, in Fig. 2(b) we show the DC susceptibility of the compound in applied field of 50 Oe. Interestingly in 50 Oe field the $T_K$ is smeared in FC branch and completely washed out in ZFC one. Because $T_K$ (35 K) occurs below $T_C$ (80 K), the competition between the two smears out the $T_K$ completely in ZFC mode at 50 Oe field. The lack of a peak at 35 K in the ZFC branch indicates, that the diamagnetic susceptibility masks the magnetic transition of small quantity TbCuO$_{2+d}$ phase. In HPLT-oxy TbSr$_2$Cu$_{2.7}$Mo$_{0.3}$O$_{7+\delta}$ compound, some characteristic X-ray lines pertaining to un-reacted phase were seen earlier [1], which could be the low dimensional magnetic phase related to $T_K$. It is important to mention that discussion given above for characteristic temperature $T_k$ is yet speculative and its real origin need to be debated.

Fig.3, depicts the ZFC magnetic susceptibility (in emu/cc units) versus applied field plot at 5 K for HPLT-oxy TbSr$_2$Cu$_{2.7}$Mo$_{0.3}$O$_7$ compound. The negative susceptibility value increases linearly with applied field of least up to 60 Oe. Considering density of the sample to be 6 gm/c.c, the estimated shielding volume fraction is around 14%. Worth mentioning is the fact at 5 K, the compound possess not only the superconductivity but a minor low dimensional magnetic phase with T$_K$ at around 35 K and the anti-ferromagnetically ordered Tb spins with T$_N$ of 7 K. In fact the HPLT-oxy TbSr$_2$Cu$_{2.7}$Mo$_{0.3}$O$_7$ compound may qualify for magneto-superconducting material.

High field (10 kOe) magnetic susceptibility of the compound above 80 K (figure not shown), exhibited Curie-Weiss paramagnetic behavior, given below,

$$\chi = \chi_0 + C/(T-\theta_P) \qquad (1)$$

Here, $\chi_0$ is the temperature independent part of the magnetic susceptibility, C is Curie-constant, related to effective paramagnetic moment of the magnetic ion, and $\theta_P$ is the Curie temperature. The values of $\chi_0$, C and $\theta_P$, being obtained after fitting the data in eq.(1) are respectively, 0.00257(2) emu/mol Oe, 10.89 ± 0.08 emu K/mol Oe and –32.9 ±0.1 K respectively. The effective paramagnetic moment of Tb in the compound being calculated from Curie constant C is 9.33 $\mu_B$.

Earlier we observed effective paramagnetic moment for Tb ions of 9.80 $\mu_B$ for non-superconducting TbSr$_2$Cu$_{2.7}$Mo$_{0.3}$O$_7$ compound [1]. In non-superconducting under-doped HTSc compounds Cu moments also contribute to magnetic susceptibility and hence the total effective moments comes out more. This further implies that in HPLT-oxy TbSr$_2$Cu$_{2.7}$Mo$_{0.3}$O$_7$ sample, the Cu is near optimally doped and has less contribution to the magnetic susceptibility. The negative value of $\theta_P$ is indicative of low temperature anti-ferromagnetic ordering of Tb spins in the compound.

Magnetization (M) versus applied field (H) plot at 5 K is shown in Fig. 4. As mentioned above, the compound has both magnetism (antiferromagnetic Tb spins with 7 K as T$_N$ and possible delafossite phase of TbCuO$_{2+d}$) and superconductivity of T$_c$ at 80 K. In such a situation the M – H plot shown in Fig.5 is supposed to have complex explanation. In recent years various high T$_c$ magneto superconductors are invented [11,12]. The M – H behavior of these compounds though theoretically explained [13], the experimental observation of the same is yet to be realized on exact theoretical model [14]. With a simplistic view of Fig.4, the full penetration field ($H_P$) is 680 Oe, the coercive field $H_c$ around 2300 Oe and remnant magnetization ($M_{rem}$) is 276 emu/mole. It is important here to note that the M-H loop shown in Fig.4 contains not only the superconducting but magnetic component also, and hence the numerical values given above can not be assigned as superconducting characteristic parameters. The M-H loops were observed right up to superconducting transition temperature of 80 K.

## SUMMARY AND CONCLUSIONS

The results of micro-structure and detailed magnetization are reported here for $TbSr_2Cu_{2.7}Mo_{0.3}O_7$ compound with superconducting transition temperature ($T_c$) of 80 K. The compound is near stoichiometric and has granular structure. Magnetization measurements at various temperatures and fields showed the compound to be bulk superconducting at 80 K. High field magnetization (10000 Oe) of the compound in normal state (> 100 K) showed Curie-Weiss behavior with a paramagnetic moment of 9.83 $\mu_B$ and a $T_N$ of 7 K for Tb moments.

## ACKNOWLEDGEMENT

This research was supported by the Klachky Foundation for Superconductivity. One of us (AVN) thanks Lady Davis Fellowship Trust for financially supporting the visit and to Prof. I. Felner for providing facilities and local hospitality.

**FIGURE CAPTIONS**

Figure 1. Typical SEM picture of HPLT-oxy TbSr$_2$Cu$_{2.7}$Mo$_{0.3}$O$_{7+\delta}$ sample.

Figure 2. DC susceptibility versus temperature plots for HPLT-oxy TbSr$_2$Cu$_{2.7}$Mo$_{0.3}$O$_{7+\delta}$ compound in both ZFC (zero-field-cooled) and FC (field-cooled) modes under an applied field of (a) 6 Oe and (b) 50 Oe.

Figure 3. ZFC magnetic susceptibility (in emu/cc units) versus applied field plot at 5 K for HPLT-oxy TbSr$_2$Cu$_{2.7}$Mo$_{0.3}$O$_7$ compound.

Figure 4. Magnetization (M) versus applied field (H) plot for HPLT-oxy TbSr$_2$Cu$_{2.7}$Mo$_{0.3}$O$_7$ compound at 5 K.

**Figure 1**

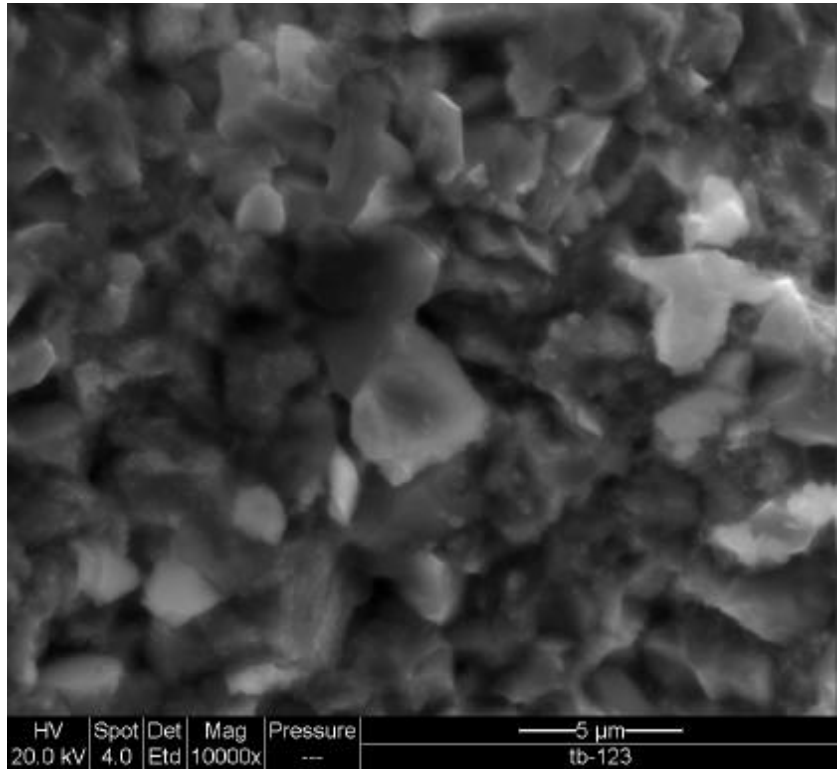

**Figure 2 (a,b)**

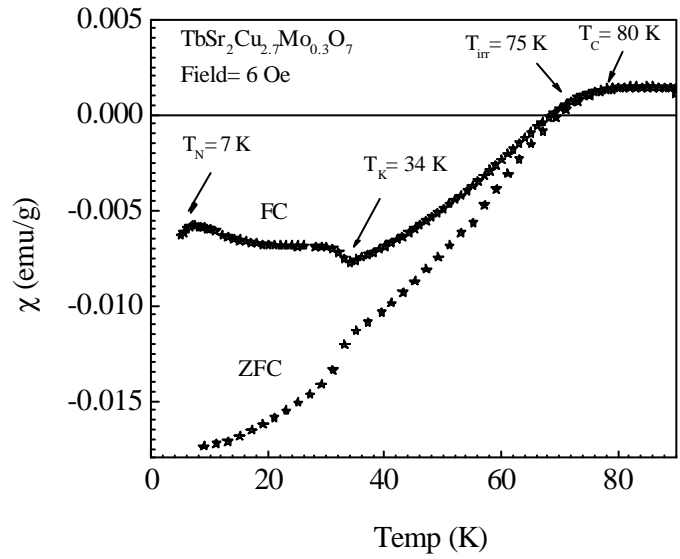

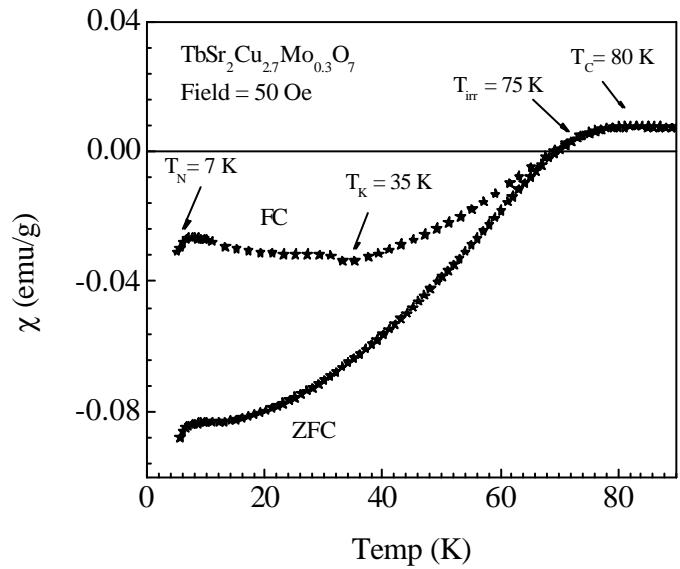

**Figure 3**

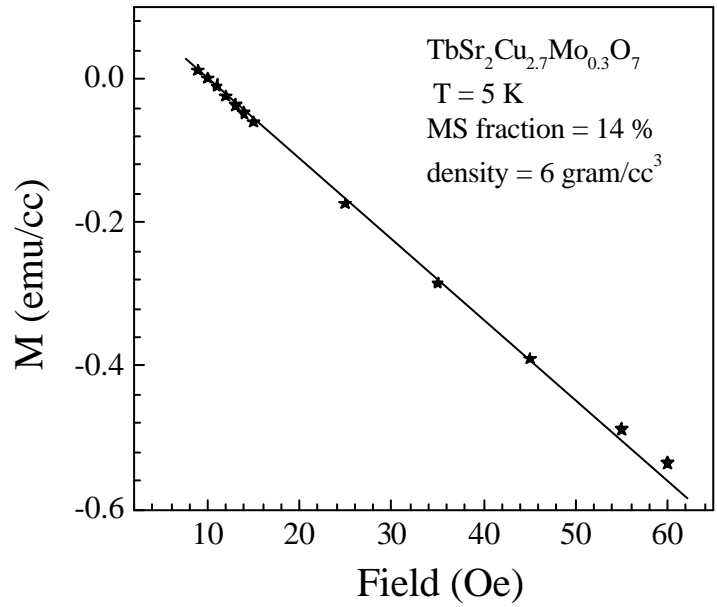

**Figure 4**

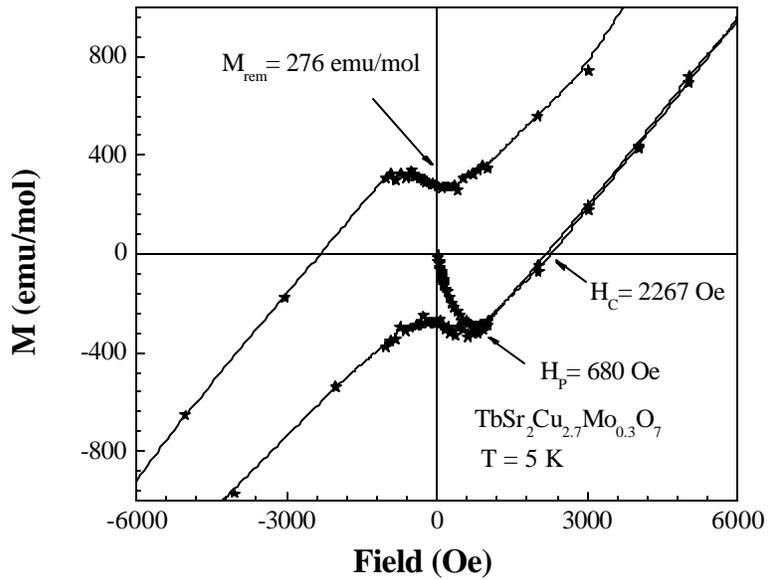